\documentclass[english,10ppt,twocolumn]{revtex4}
\usepackage[T1]{fontenc}
\usepackage[latin9]{inputenc}
\usepackage{amsmath}
\usepackage{amssymb}
\usepackage{graphicx}
\usepackage{color}

\makeatletter
\@ifundefined{textcolor}{}
{%
 \definecolor{BLACK}{gray}{0}
 \definecolor{WHITE}{gray}{1}
 \definecolor{RED}{rgb}{1,0,0}
 \definecolor{GREEN}{rgb}{0,1,0}
 \definecolor{BLUE}{rgb}{0,0,1}
 \definecolor{CYAN}{cmyk}{1,0,0,0}
 \definecolor{MAGENTA}{cmyk}{0,1,0,0}
 \definecolor{YELLOW}{cmyk}{0,0,1,0}
}

\usepackage{babel}

\makeatother

\usepackage{babel}
\begin{document}

\preprint{This line only printed with preprint option}

\title{de Broglie-Proca and Bopp-Podolsky massive photon gases in cosmology}

\author{R.R. Cuzinatto}
\email{rodrigo.cuzinatto@unifal-mg.edu.br}

\affiliation{Department of Physics, McGill University, Ernest Rutherford Physics
Building, 3600 University Street, H3A 2T8, Montreal, QB, Canada}

\address{Instituto de Ci\^encia e Tecnologia, Universidade Federal de Alfenas,
Rodovia Jos\'e Aur\'elio Vilela, 11999, Cidade Universit\'aria, CEP 37715-400,
Po\c cos de Caldas, MG, Brazil}

\author{E.M. de Morais}
\email{eduardomessiasdemorais@gmail.com }

\address{Instituto de Ci\^encia e Tecnologia, Universidade Federal de Alfenas,
Rodovia Jos\'e Aur\'elio Vilela, 11999, Cidade Universit\'aria, CEP 37715-400,
Po\c cos de Caldas, MG, Brazil}

\author{L.G. Medeiros}
\email{leogmedeiros@ect.ufrn.br}

\address{Escola de Ci\^encia e Tecnologia, Universidade Federal do Rio Grande
do Norte, Campus Universit\'ario, s/n, CEP 59072-970, Natal, Brazil}

\address{Instituto de F\'isica Te\'orica, S\~ao Paulo State University, P.O. Box
70532-2, CEP 01156-970 S\~ao Paulo, SP, Brazil}

\author{C. Naldoni de Souza}
\email{clicia1.2009@gmail.com }

\address{Instituto de Ci\^encia e Tecnologia, Universidade Federal de Alfenas,
Rodovia Jos\'e Aur\'elio Vilela, 11999, Cidade Universit\'aria, CEP 37715-400,
Po\c cos de Caldas, MG, Brazil}

\author{B.M. Pimentel}
\email{pimentel@ift.unesp.br}

\address{Instituto de F\'isica Te\'orica, S\~ao Paulo State University, P.O. Box
70532-2, CEP 01156-970 S\~ao Paulo, SP, Brazil}

\begin{abstract}
We investigate the influence of massive photons on the evolution of
the expanding universe. Two particular
models for generalized electrodynamics are considered, namely de Broglie-Proca and Bopp-Podolsky electrodynamics. We obtain the equation of state
(EOS) $P=P(\varepsilon)$ for each case using dispersion relations
derived from both theories. The EOS are inputted into the Friedmann
equations of a homogeneous and isotropic space-time to determine the
cosmic scale factor $a(t)$. It is shown that the photon non-null
mass does not significantly alter the result $a\propto t^{1/2}$ valid
for a massless photon gas; this is true either in de Broglie-Proca's case (where
the photon mass $m$ is extremely small) or in Bopp-Podolsky theory (for
which $m$ is extremely large).
\end{abstract}
\maketitle

\section{Introduction\label{sec:Introduction}}

Physical cosmology assumes a homogeneous and isotropic universe in
very large scales \cite{Ryden}. These symmetry requirements lead
to major simplifications on Einstein's equations of general relativity
\cite{DeSabbata}, which reduce to the so-called Friedmann and conservation
equations
\begin{eqnarray}
\left(\frac{\dot{a}}{a}\right)^{2} & = & \frac{8\pi G}{3}\varepsilon\,,\label{eq:FriedEq}
\end{eqnarray}
\begin{equation}
\frac{d\varepsilon}{da}+\frac{3}{a}(P+\varepsilon)=0\,,\label{eq:FluidEq}
\end{equation}
where $a=a(t)$ is the scale factor, a function of cosmic time $t$
related to distances in the cosmos. (The dot on top of variables denotes a
time derivative.) We are neglecting the cosmological constant ($\Lambda=0$),
the spatial section of space-time is taken as flat (the curvature
parameter is taken as null, $\kappa=0$) and $G$ stands for Newtonian
gravitational constant. $\varepsilon$ is the energy density
associated with the matter-energy content assumed to fill the universe.

Baryonic matter is usually described as an incoherent set of particles
respecting the dust-like equation of state (EOS), with null pressure: $P=0$.
Radiation is treated as a thermalized massless photon gas in accordance
with Maxwell electrodynamics; then, blackbody statistical mechanics \cite{Pathria2011} gives $P=\varepsilon/3$ for the EOS of the radiation content. Substitution of these two EOS into (\ref{eq:FluidEq})
leads to $\varepsilon\propto a^{-3}$ and $\varepsilon\propto a^{-4}$
for matter and radiation respectively. Inserting these formulas of
$\varepsilon=\varepsilon(a)$ into (\ref{eq:FriedEq}) results the
dynamics $a\propto t^{2/3}$ for dust and $a\propto t^{1/2}$ in the
case of radiation. This means in an expanding universe, the contribution
from radiation is energetically more relevant in the early universe
whereas baryonic matter is comparatively more important to cosmic
dynamics (i.e. the time evolution of the scale factor) at later times. One might ask how this whole picture would change if, instead of being massless as in Maxwell electrodynamics,
the photon had a mass. The present paper is an attempt to address
this point.

Naturally, the relevance of this question is deeply connected to the
importance one gives alternatives to the standard theory of electromagnetism. Maxwell's theory has been remarkably well tested through a plethora of experiments and observations \cite{Tu,Goldhaber2010}. Modifications to Maxwellian electromagnetism, such as
de Broglie-Proca \cite{deBroglie1922,deBroglie1923,deBroglie,Proca} and Bopp-Podolsky \cite{Bopp,Podolsky,Lande} theories, introduce a non-null mass for the photon \footnote{The approach by Bopp and Podolsky is based on modifying the ordinary Lagrangian of electrodynamics. Land\'e contribution had a different motivation -- namely to address the problem of electron self energy -- but he himself soon realized the equivalence between his proposal and the one by Bopp.}. Should this mass have consequences for cosmic dynamics which are detectable, then cosmological observations
could be an instrument to set constraints on the value of the photon
mass and, at the same time, serve as a testing ground for the standard
and alternative theories of electromagnetism.

de Broglie-Proca field equations are the simplest relativistic way to introduce
mass in electromagnetism \cite{Tu} since the vector potential $A_{\mu}(x)$
respects a Klein-Gordon equation; moreover, the Wentzel-Pauli Lagrangian
\cite{Aldrovandi,B-Field} leading to de Broglie-Proca electromagnetism presents
no additional derivative terms on $A_{\mu}$ besides those making
up the field strength $F^{\mu\nu}$ \textendash{} see Sect.  ``de Broglie-Proca cosmology'' below.
Experimental constraints on the mass of the de Broglie-Proca photon are very restrictive;
they are given in \cite{Tu,Goldhaber2010,Ryutov2007,PDG2016,Bonetti2016} and demand it to be extremely
small. The Stueckelberg field \cite{Stueckelberg,Ruegg} does not
bear higher-order derivative terms in its field equations and has
the additional feature of preserving gauge invariance; however one
pays the price of introducing an extra scalar field $B(x)$. Generalizations
of de Broglie-Proca's and Stueckelberg's approaches are available today; see e.g.
\cite{Allys} and references therein.

Podolsky's Generalized Electrodynamics \cite{Podolsky,PodolskySchwed}
differs from the previous cases by exhibiting derivative couplings.
Bopp-Podolsky action includes derivatives of $F^{\mu\nu}$, a fact that
leads to field equations for the vector potential with order higher
than two \textendash{} cf. Sect. ``Bopp-Podolsky cosmology''. These
additional terms were introduced to make the resulting generalized
quantum electrodynamics (GQED) regular in the first order \cite{PodolskySchwed}.
Moreover, in Bopp-Podolsky the extra term generates a massive mode which preserves
the $U(1)$ gauge invariance without the necessity of introducing
new fields. Literature offers references with classical \cite{AcciolyMukai}
and quantum \cite{Bufalo2011,Bufalo2013,Bufalo2014} developments
of Bopp-Podolsky's proposal; some of those works impose bounds on the massive
Bopp-Podolsky photon \cite{ProbePodolsky,Bonin2009,Bufalo2012}.

The above generalizations of Maxwell electromagnetism may be classified
as linear theories. Conversely,
there are non-linear electrodynamics (NLED) \cite{Plebanski1968}
 coming from Euler-Heisenberg \cite{EH1936,Holger2000}
and Born-Infeld \cite{BI1934a,BI1934b} Lagrangians. NLED are a clear
example of how important modifications to Maxwell electromagnetism
can be to cosmology: they may offer an explanation to accelerating
universe \cite{NoveloSalim,Krulov}, generate bouncing \cite{NovelloSalimLorenci} and
produce cyclic universes \cite{CyclicUniverse,Leo2012}.

Some attempts have been made to investigate the influence of massive
photons in cosmology in the context of the generalized Proca electrodynamics
\cite{MassPhoDE,EleCosmScal,CosmGenProca} and Bopp-Podolsky theory \cite{Haghani};
however, these approaches were implemented via field theory. As far
as these authors are aware, none of the mentioned works address this
problem through a thermodynamical approach, using EOS
built from the statistical treatment of the massive photon gas. That
is what we  perform in the following sections. %Sect. \ref{sec:Proca-Cosmology}
%deals with the possible consequences of a light photon to cosmic dynamics,
%the Proca case. Sect. \ref{sec:Podolsky-Cosmology} presents the EOS
%for Podolsky radiation and uses it in the Friedmann equations to determine
%how the mass affects the thermal history of the universe (if it has
%any influence whatsoever). In Sect. \ref{sec:final-remarks} we summarize
%our results and give our interpretations and conclusions.

\section{de Broglie-Proca cosmology\label{sec:Proca-Cosmology}}

The Lagrangian of de Broglie-Proca electrodynamics in vacuum is:
\begin{equation}
{\cal L}=-\frac{1}{4}F^{\mu\nu}F_{\mu\nu}+\frac{1}{2}m^{2}A^{\mu}A_{\mu}\,,\label{eq:L_Proca}
\end{equation}
where
\begin{equation}
F_{\mu\nu}=\partial_{\mu}A_{\nu}-\partial_{\nu}A_{\mu}\,.\label{eq:Fmunu}
\end{equation}
The massive term in Eq. (\ref{eq:L_Proca}) violates gauge invariance
which makes it arguably the introduction of the field strength (\ref{eq:Fmunu})
in a deductive way as done in \cite{Utyiama}. Nevertheless, the photon
mass $m$ is admittedly small so that de Broglie-Proca term is a correction
to Maxwell's theory; in fact, experimental constraints set \cite{Ryutov2007,PDG2016} \footnote{Reference \cite{Retino2016} shows that the result in \cite{Ryutov2007} is partly speculative. However, even if the constraint is as high as $m \lesssim 10^{-13}\mbox{eV}$ the conclusions presented here would essentially remain the same.}:
\begin{equation}
m\leq10^{-18}\mbox{eV}\qquad(\mbox{de Broglie-Proca})\, . \label{eq:ProcaMass}
\end{equation}

From the de Broglie-Proca Lagrangian we obtain the following vacuum field equations:
\begin{equation}
\partial_{\mu}F^{\mu\nu}+m^{2}A^{\nu}=0\,.\label{eq:FieldEqProca}
\end{equation}
Applying $\partial_{\nu}$ to Eq. (\ref{eq:FieldEqProca}) and using
the antisymmetry property of $F^{\mu\nu}$, one checks that de Broglie-Proca field
satisfies
\begin{equation}
\partial_{\mu}A^{\mu}=0\,,\label{eq:LorenzCondition}
\end{equation}
which is the ordinary Lorenz condition. This relation is a constraint
reducing the degrees of freedom of the theory to three. 

Using (\ref{eq:LorenzCondition}), the equations of motion
(\ref{eq:FieldEqProca}) may be written in terms of the potential $A^{\mu}$:
\begin{equation}
\left(\square +m^{2}\right) A^{\mu}=0\,.\label{eq:EqMotionProcaA}
\end{equation}
Then, by means of a Fourier transform,
\begin{equation}
A^{\mu}(x)=\frac{1}{(2\pi)^{4}}\int\bar{A}^{\mu}(k)e^{-ik_{\nu}x^{\nu}}d^{4}x\,,\label{eq:FourierTrans}
\end{equation}
one obtains the dispersion relation
\begin{equation}
k_{\mu}k^{\mu}=m^{2}\Rightarrow\omega^{2}=m^{2}+{\bf p}^{2}\, , \label{eq:RelativisticEnergy}
\end{equation}
where $k^{0}=\omega$ and $p^{i}=k^{i}$ in units where $c=\hbar=1$.

The de Broglie-Proca field is a vector boson. Due to this nature, the canonical
partition function associated with the massive photon gas is \cite{Pathria2011}:
\begin{align}
\ln Z & =-\frac{g}{(2\pi)^{3}}\int d^{3}{\bf x}\int d^{3}{\bf p}\ln\left(1-e^{-\beta\omega}\right) \nonumber \\
 & =-\frac{g}{2}\frac{m^{2}}{\pi^{2}}V\sum_{k=1}^{\infty}\frac{K_{2}(k\beta m)}{k^{2}\beta}\,,\label{eq:PartFuncProca}
\end{align}
where $g$ is the number of internal degrees of freedom, parameter
$\beta=\frac{1}{T}$ is the inverse of the temperature $T$ (in units
of normalized Boltzmann constant, $k_{B}=1$), $K_{2}$ is the modified
Bessel function of the second kind \cite{Gradshteyn} and $V$ is
the volume occupied by the gas.

The partition function $Z(\beta,V;m)$ is a key ingredient for obtaining
the energy density $\varepsilon$ and the pressure $P$ of the massive
photon gas \cite{Pathria2011}:
\begin{equation}
\varepsilon=-\frac{1}{V}\frac{\partial}{\partial\beta}\ln Z\text{ \ \ and \ \ }P=\frac{1}{\beta}\frac{\partial}{\partial V}\ln Z\,.\label{eq:EnergyPressure}
\end{equation}

%In principle, the three degrees of freedom admitted by Proca's theory
%would demand $g=3$. Those degrees correspond to the propagation modes
%of the photon: There are two transversal modes and one longitudinal
%propagation mode. However, the longitudinal mode interacts weakly
%with matter \cite{Tu}. Thus, the oscillators on the walls of the
%blackbody's cavity interact significantly only with the two transversal
%modes of the massive photon, leading to thermal equilibrium through
%a statistical process with $g=2$. According to this reasoning, one
%should take $g=2$ for the Proca field. Nevertheless, in what follows
%we shall not use the numerical value for $g$ since it is not relevant
%for our conclusions.

By substituting (\ref{eq:PartFuncProca}) into (\ref{eq:EnergyPressure}),
one calculates:
\begin{equation}
P=\frac{g}{2}\frac{m^{4}}{\pi^{2}}\sum_{k=1}^{\infty}\frac{K_{2}(k\beta m)}{\left(k\beta m\right)^{2}}\,,\label{eq:PressureProca}
\end{equation}
and
\begin{equation}
\varepsilon-3P=\frac{g}{2}\frac{m^{4}}{\pi^{2}}\sum_{k=1}^{\infty}\frac{K_{1}(k\beta m)}{\left(k\beta m\right)}\,.\label{eq:EnergyProca}
\end{equation}
Notice that $K_{1}(z)\simeq z^{-1}$ for $z\ll1$. Therefore Eq. (\ref{eq:EnergyProca})
leads to $P=\frac{1}{3}\varepsilon$ in the limit as $m\rightarrow0$,
which is the expected result for the blackbody radiation of a massless
photon gas as in Maxwell electrodynamics.

Let us now turn to the study of the cosmic dynamics for a de Broglie-Proca photon
gas. 

In order to solve Eq. (\ref{eq:FluidEq}) one requires an equation
of state. However, it is clear that we can not analytically invert
Eq. (\ref{eq:EnergyProca}) for obtaining $\beta=\beta(\varepsilon)$,
which would be, in turn, substituted into (\ref{eq:PressureProca})
leading to $P=P(\varepsilon)$. Therefore, there is no analytical
function $\varepsilon=\varepsilon(a)$ to be inserted into Friedmann
equation (\ref{eq:FriedEq}) which would be integrated to give $a=a(t)$.
Of course, we could solve the pair of equations for cosmology
(\ref{eq:FriedEq}-\ref{eq:FluidEq}) along with the constitutive
equations (\ref{eq:PressureProca}-\ref{eq:EnergyProca}) for de Broglie-Proca
electrodynamics numerically. Nevertheless, it is possible to obtain
approximated analytic solutions in the limits as $\beta m\ll1$ or
$\beta m\gg1$ which are physically meaningful.

The property
\begin{equation}
K_{2}(z)\simeq\frac{2}{z^{2}}-\frac{1}{2}\qquad(z\ll1)\label{eq:K2zSmall}
\end{equation}
is useful to analyze the limit $\beta m\ll1$. In this case, pressure
and energy density for a de Broglie-Proca photon gas assume the following simple
forms:
\begin{equation}
P\simeq\frac{1}{3}\left(\frac{\pi^{2}}{15}\frac{1}{\beta^{4}}\right)\frac{g}{2}\left(1-3\frac{5}{4}\frac{1}{\pi^{2}}\left(\beta m\right)^{2}\right) \,\,\, (\beta m\ll1)\,,\label{eq:PProcaBetaMSmall}
\end{equation}
\begin{equation}
\varepsilon\simeq\left(\frac{\pi^{2}}{15}\frac{1}{\beta^{4}}\right)\frac{g}{2}\left(1-\frac{5}{4}\frac{1}{\pi^{2}}\left(\beta m\right)^{2}\right) \qquad (\beta m\ll1)\,.\label{eq:EProcaBetaMSmall}
\end{equation}
Eq. (\ref{eq:EProcaBetaMSmall}) can be promptly inverted and substituted
into (\ref{eq:PProcaBetaMSmall}) to give:
\begin{equation}
P\simeq\frac{\varepsilon}{3}\left(1-4\frac{M^{2}}{\sqrt{\varepsilon}}\right)\qquad(\beta m\ll1)\,,\label{eq:P(E)ProcaBetaMSmall}
\end{equation}
where
\begin{equation}
M^{2}=M^{2}(m)\equiv\frac{1}{6}\sqrt{\frac{15}{\pi^{2}}}\sqrt{\frac{g}{2}}\frac{m^{2}}{4}\qquad(\mbox{de Broglie-Proca})\,.\label{eq:MProca}
\end{equation}
It is worth noting that $\frac{M^{2}}{\sqrt{\varepsilon}}\ll1$ since
$\beta m\ll1$. By substituting (\ref{eq:P(E)ProcaBetaMSmall}) into
(\ref{eq:FluidEq}), it results in:
\begin{equation}
\frac{d\varepsilon}{da}+\frac{4}{a}\varepsilon\left(1-\frac{M^{2}}{\sqrt{\varepsilon}}\right)=0\,,\label{eq:DiffEqE(a)}
\end{equation}
which is immediately integrated to give:
\begin{equation}
\varepsilon(a)=\varepsilon_{0}\left(\frac{a_{0}}{a}\right)^{4}\left[1-\frac{M^{2}}{\sqrt{\varepsilon_{0}}}\left(1-\frac{a^{2}}{a_{0}^{2}}\right)\right]^{2}\qquad(\beta m\ll1)\,,\label{eq:E(a)ProcaBetaMSmall}
\end{equation}
under the integration condition $\varepsilon(a_{0})=\varepsilon_{0}$
and taking $a_{0}$ as an arbitrary fixed value of the scale factor, such as its present-day value. Eq.~(\ref{eq:E(a)ProcaBetaMSmall})
is the same as expected for a radiation gas in standard cosmology
plus a correction due to the (small) value of the de Broglie-Proca mass. The last
step in the cosmological analysis is to substitute (\ref{eq:E(a)ProcaBetaMSmall})
in the Friedmann equation (\ref{eq:FriedEq}) and integrate the resulting
differential equation. This leads to:
\begin{equation}
a \simeq a_{0} \left\{ 1+2H_{0}(t-t_{0})\left[1+\frac{M^{2}}{\sqrt{\varepsilon_{0}}}H_{0}(t-t_{0})\right]\right\}^{1/2} \label{eq:a(t)ProcaBetaMSmall}
\end{equation}
$(\beta m\ll1)$, with the initial condition $a(t_{0})=a_{0}$ and $H_{0}=\sqrt{8\pi G\varepsilon_{0}/3}$
is the Hubble function $H=\dot{a}/a$ calculated at the time $t=t_{0}$.
Solution (\ref{eq:a(t)ProcaBetaMSmall}) is precisely the scale factor
for the standard radiation era plus a small extra term which depends
on $m$.

The maximum possible mass value for the photon in de Broglie-Proca theory
allowed by experimental constraints is $m=10^{-18}\mbox{eV}$, cf.~\cite{PDG2016}. This means that the condition $\beta m\ll1$ is consistent
with temperatures ranging from extremely high values till values of
the order $T_{P} \sim 10^{-18}\mbox{eV} \sim 10^{-14}\mbox{K}$,
corresponding to the distant future universe. In fact, the temperature
and the scale factor are roughly inversely proportional, so that
\begin{equation}
\frac{a_{P}}{a_{0}} \sim \frac{T_{0}}{T_{P}}\Rightarrow a_{P} \sim 10^{14}a_{0}\,.\label{eq:aP}
\end{equation}
is the estimate for the scale factor above which the influence of
de Broglie-Proca mass in cosmology is appreciable. ($T_{0}\simeq2.73\mbox{K}$
is the cosmic microwave background radiation temperature today.) The
bottom line is that the condition $\beta m\ll1$ applies whenever
$a\ll10^{14}a_{0}$, i.e., for all values of $a$ less than $10^{14}$ the present day scale factor, the influence of de Broglie-Proca mass in cosmic dynamics is negligible: this encompasses all the period from
the primeval universe up to the present and towards the distant future.
This conclusion is confirmed by the study of the theory in the other
limit for $\beta m$, below.

For the limit $\beta m\gg1$, the convenient asymptotic form for the
modified Bessel functions is:
\begin{equation}
K_{1}(z)\simeq K_{2}(z)\simeq\sqrt{\frac{\pi}{2z}}e^{-z}\qquad(z\gg1)\,.\label{eq:K2zBig}
\end{equation}
By substituting this result in Eqs. (\ref{eq:PressureProca}) and
(\ref{eq:EnergyProca}) and keeping only the first term in the sums
over index $k$, one gets:
\begin{equation}
P\simeq\frac{g}{2}\frac{1}{\beta^{4}}\frac{(\beta m)^{3/2}}{\sqrt{2\pi^{3}}}e^{-\beta m}\qquad(\beta m\gg1)\,,\label{eq:PProcaBetaMBig}
\end{equation}
\begin{equation}
\varepsilon\simeq\frac{g}{2}\frac{1}{\beta^{4}}\frac{(\beta m)^{5/2}}{\sqrt{2\pi^{3}}}e^{-\beta m}\qquad(\beta m\gg1)\,,\label{eq:EProcaBetaMBig}
\end{equation}
so that the equation of state is:
\begin{equation}
\frac{P}{\varepsilon}\simeq\frac{1}{\beta m}\qquad(\beta m\gg1)\,,\label{eq:P(E)ProcaBetaMBig}
\end{equation}
Therefore, $P\ll\varepsilon$ in the limit $\beta m\gg1$ and one
can adopt the dust approximation for incoherent particles: $P\simeq0$.
As a consequence, Eqs. (\ref{eq:FriedEq}, \ref{eq:FluidEq}) lead
to:
\begin{equation}
\varepsilon=\varepsilon_{0}\left(\frac{a_{0}}{a}\right)^{3}\text{ \ \ and \ \ }a\sim t^{2/3}\qquad(\beta m\gg1)\,,\label{eq:a(t)ProcaBetaMBig}
\end{equation}
which are the equations for non-relativistic matter in cosmology.

Notice that the condition $\beta m\gg1$ is violated for values of
$\beta$ which can not compensate the extremely small value of $m$.
Hence, the condition is consistent with high values of $\beta$, or
conversely small values of $T$, namely $T\sim T_{P}$. Thus, in the
limit $\beta m\gg1$ we are dealing with the distant future universe,
far larger than $a_{P}$.

From all the discussion above, we notice that the energy density of
the massive photon in de Broglie-Proca theory is either practically the same as
the massless photon of Maxwell theory ($\varepsilon\sim a^{-4}$)
or it scales as the energy density of ordinary and dark matter ($\varepsilon\sim a^{-3}$).
On the other hand, baryonic and dark matter are much more abundant
than radiation today. Therefore the influence of de Broglie-Proca electrodynamics
is negligible for the cosmic dynamics.

%%%%%%%%%%%%%%%%%%%%%%%%%%%%%%%%%%%%%%%%%%%%%%%%

\section{Bopp-Podolsky cosmology\label{sec:Podolsky-Cosmology}}

Podolsky's Generalized Electrodynamics is derived from the Lagrangian
\begin{equation}
{\cal L}=-\frac{1}{4}F_{\mu\nu}F^{\mu\nu}+\frac{a^{2}}{2}\partial_{\mu}F^{\mu\nu}\partial_{\rho}F_{\,\nu}^{\rho}\,,\label{eq:L_Podolsky}
\end{equation}
where the field strength $F^{\mu\nu}$ is defined in (\ref{eq:Fmunu}).
Unlike de Broglie-Proca's case, this theory is completely consistent with Utiyama's
procedure for building a gauge theory from a symmetry requirement
\cite{2ndGaugeTheory}.

The field equation in the absence of sources is:
\begin{equation}
\left(1+a^{2}\square\right)\partial_{\mu}F^{\mu\nu}=0\,.\label{eq:FieldEqPodolsky}
\end{equation}
When one writes (\ref{eq:FieldEqPodolsky}) in terms of $A^{\mu}$
and uses the generalized Lorenz gauge condition \cite{PimentelGalvao}
\begin{equation}
\left(1+a^{2}\square\right)\partial_{\mu}A^{\mu}=0\,,\label{eq:GeneralLorenzCondition}
\end{equation}
it results in
\begin{equation}
\left(1+a^{2}\square\right)\square A^{\mu}=0\,.\label{eq:EqMotion}
\end{equation}
The r.h.s. of this equation equals the four-current $j^{\mu}$ if
there are sources. Using (\ref{eq:FourierTrans}), Eq. (\ref{eq:EqMotion})
implies two independent dispersion relations for Bopp-Podolsky photon:
\begin{equation}
k^{\mu}k_{\mu}=0\Rightarrow\omega^{2}=p^{2}\,,\label{eq:EnergyMax}
\end{equation}
and
\begin{equation}
1-a^{2}k^{\mu}k_{\mu}=0\Rightarrow\omega^{2}=\frac{1}{a^{2}}+{\bf p}^{2}\,.\label{eq:EnergyPod}
\end{equation}
The first dispersion relation is the one typical of a massless photon
and the second one is the same as (\ref{eq:RelativisticEnergy}) under
the identification
\begin{equation}
m^{2}=\frac{1}{a^{2}}\,.\label{eq:MassPod}
\end{equation}
Eqs.~(\ref{eq:EnergyPod}-\ref{eq:MassPod}) are the reason for attributing
a non-zero mass to Bopp-Podolsky photon. In fact, if the Bopp-Podolsky term in the
Lagrangian is supposed to represent only a correction to Maxwell electrodynamics,
then the coupling constant $a$ should be very small, i.e., the Bopp-Podolsky
photon mass $m$ should be very large. Consequently, one might expect
a Bopp-Podolsky-type photon gas to be relevant for the cosmic dynamics
of the early universe, when the mean energy is high enough to access
the massive mode for the photon. One of the goals of this paper is
to check this hypothesis.

Eqs.~(\ref{eq:EnergyMax}) and (\ref{eq:EnergyPod}) also show the
separation of Bopp-Podolsky theory into a massless mode (Maxwell) and massive
mode (a de Broglie-Proca-type dispersion relation except for the hugeness of
the mass). As a consequence, the partition function for a Bopp-Podolsky photon
gas will bare two terms, each one related to a different mode:
\begin{align}
\ln Z= & -\frac{g_{M}V}{\left(2\pi\right)^{3}}\int d^{3}{\bf p}\ln\left(1-e^{-\beta{\bf p}}\right) \nonumber \\
& -\frac{gV}{\left(2\pi\right)^{3}}\int d^{3}{\bf p}\ln\left(1-e^{-\beta \sqrt{{\bf p}^{2}+m^{2}}}\right)\,.\label{eq:PartFuncPodolsky}
\end{align}
The first term of the r.h.s. is the ordinary partition function for
the massless photon of Maxwell electrodynamics with helicity two,
meaning $g_{M}=2$ for the number of internal degrees of freedom.
The second term of the r.h.s. of Eq. (\ref{eq:PartFuncPodolsky})
is the de Broglie-Proca-like contribution $-$ compare with Eq. (\ref{eq:PartFuncProca}).
In spite of the presence of a de Broglie-Proca-like term in the thermodynamics
of Bopp-Podolsky massive photon gas, one should not expect the same consequences
derived in the previous section to hold here. There
is a crucial difference concerning the photon mass: for Bopp-Podolsky's
case $m\gg1$, whilst in de Broglie-Proca's case $m\ll1$. Moreover, the assumption
$g=3$ for the massive sector of Bopp-Podolsky theory is consistent
with blackbody radiation measurements.

The first integral in Eq. (\ref{eq:PartFuncPodolsky}) is found in
standard text-books on statistical mechanics, see, e.g., \cite{Pathria2011};
the second integral was solved in Sect. ``de Broglie-Proca cosmology''.
Hence, Bopp-Podolsky partition function is:
\begin{equation}
\ln Z=\frac{\pi^{2}}{45}\frac{V}{\beta^{3}}+\frac{g}{2}\frac{m^{3}}{\pi^{2}}V\left(\beta m\right)\sum_{k=1}^{\infty}\frac{K_{2}(k\beta m)}{\left(k\beta m\right)^{2}}\,.\label{eq:PartFuncPodolskyK2}
\end{equation}
Eqs. (\ref{eq:EnergyPressure}) and (\ref{eq:PartFuncPodolskyK2})
lead to:
\begin{equation}
P=\frac{\pi^{2}}{45\beta^{4}}\left[1+45\frac{g}{2}\frac{\left(\beta m\right)^{4}}{\pi^{4}}\sum_{k=1}^{\infty}\frac{K_{2}(k\beta m)}{\left(k\beta m\right)^{2}}\right]\,,\label{eq:PressurePodolsky}
\end{equation}
and
\begin{equation}
\varepsilon-3P=\frac{g}{2}\frac{m^{4}}{\pi^{2}}\sum_{k=1}^{\infty}\frac{K_{1}(k\beta m)}{\left(k\beta m\right)}\,.\label{eq:EnergyPodolsky}
\end{equation}
Notice that Eqs. (\ref{eq:EnergyProca}) and (\ref{eq:EnergyPodolsky})
for the energy density of de Broglie-Proca and Bopp-Podolsky theories are formally
the same. However, the values for $\varepsilon(\beta,m)$ will not
be the same since the pressures in Eqs. (\ref{eq:PressureProca})
and (\ref{eq:PressurePodolsky}) are different.

Eq. (\ref{eq:EnergyPodolsky}) may be written as:
\begin{equation}
\varepsilon=\varepsilon_{M}\left(1+\delta\varepsilon\right)\,,\label{eq:EnergyPodolskyDeltaE}
\end{equation}
where
\begin{equation}
\varepsilon_{M}=\frac{\pi^{2}}{15}\frac{1}{\beta^{4}}\label{eq:EnergyMaxwell}
\end{equation}
is the energy density of a Maxwellian massless photon gas and
\begin{equation}
\delta\varepsilon=15\frac{g}{2}\frac{\left(\beta m\right)^{4}}{\pi^{4}}\sum_{k=1}^{\infty}\left[\frac{K_{1}(k\beta m)}{\left(k\beta m\right)}+3\frac{K_{2}(k\beta m)}{\left(k\beta m\right)^{2}}\right]\label{eq:DeltaE}
\end{equation}
is the correction due to Bopp-Podolsky mass.

%%% Figure %%%

\begin{figure}[ht]
\begin{centering}
\includegraphics[height=6.4cm, width=8.5cm]{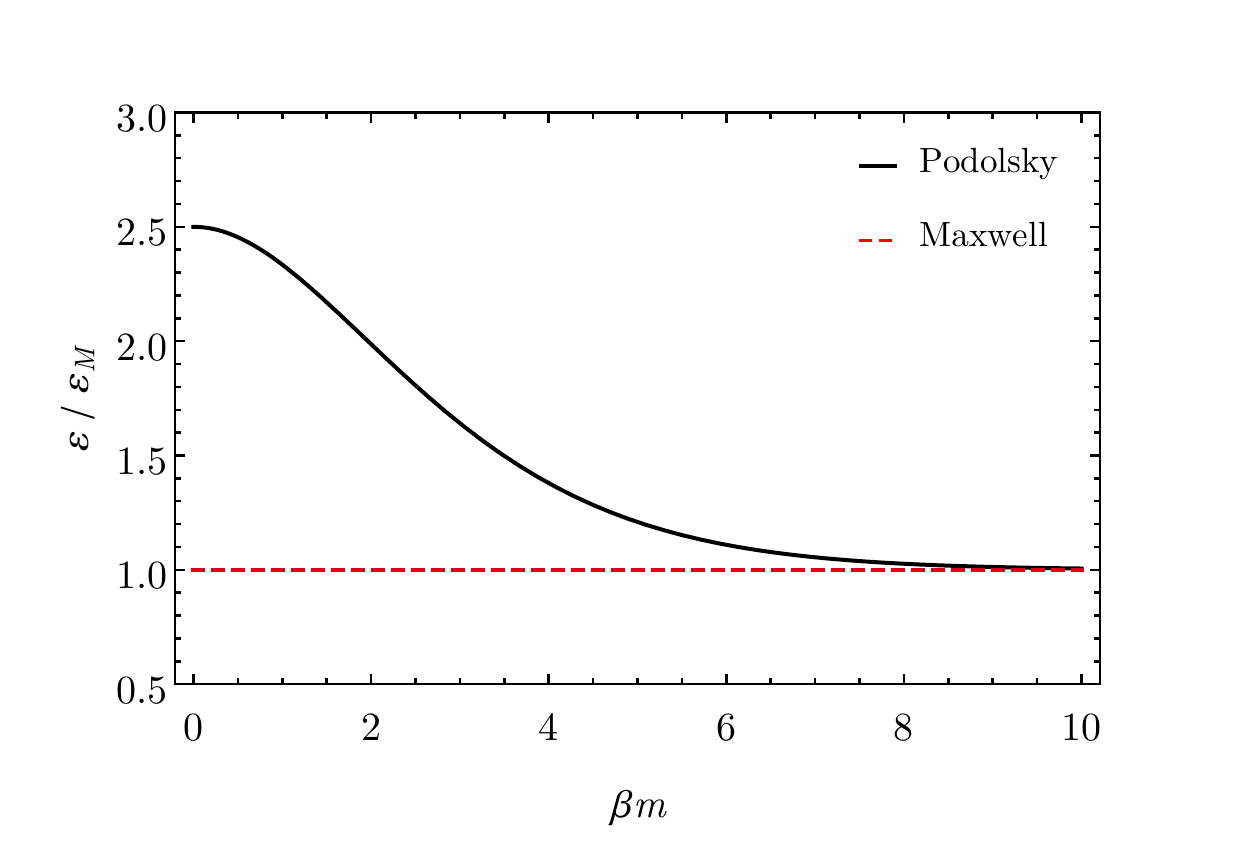}
\par\end{centering}
\caption{Plot of $\varepsilon/\varepsilon_{M}$ as a function of parameter
$\beta m$. It is assumed that the degeneracy degree of Bopp-Podolsky radiation
is $g=3$. The energy density of Bopp-Podolsky massive photon gas tends
to the ordinary energy density of a massless photon gas as $\beta m$
assumes values larger than $\sim$10.}
\label{fig:E(BetaM)}
\end{figure}

%%%%%%%%%%

Fig. \ref{fig:E(BetaM)} shows the plot of $\varepsilon/\varepsilon_{M}$
as a function of the dimensionless parameter $\beta m$. It is assumed
$g=3$ for Bopp-Podolsky photon gas. One notices that $\delta\varepsilon$
approaches $g/2$ as $\beta m$ approaches zero, so that $\lim_{\beta m\rightarrow0}\left(\varepsilon/\varepsilon_{M}\right)=2.5$
$-$ see Eqs. (\ref{eq:EnergyPodolskyDeltaE}) and (\ref{eq:DeltaE}).
The same plot also shows that $\delta\varepsilon$ is negligible for
large values of $\beta m$ once the curve for $\varepsilon/\varepsilon_{M}$
approaches $1$ from $\beta m\sim10$ (a condition that is
guaranteed for a temperature ten times smaller than the rest mass
of the Bopp-Podolsky photon). Ref. \cite{Bufalo2012} sets the most restrictive
limit for the mass of the photon in Podolsky Generalized Electrodynamics
known today, namely
\begin{equation}
m\gtrsim3.7\times10^{10}\mbox{eV}\gtrsim10^{14}\mbox{K}\qquad(\mbox{Bopp-Podolsky})\,;\label{eq:PodolskyMass}
\end{equation}
this is the scale of energy where one expects $\delta\varepsilon$
being relevant. This energy scale corresponds to the early universe,
way before the quark-gluon deconfinement. The primeval universe is consistent with
the regime where $\beta m\ll1$ in Bopp-Podolsky theory. This limit and
its implication to cosmology are analyzed below.

Eq.~(\ref{eq:PartFuncPodolskyK2}) may be simplified in the limit
$\beta m\ll1$. The resulting expression for $\ln Z(V,\beta;m)$ is
then substituted into Eq. (\ref{eq:EnergyPressure}) yielding:
\begin{equation}
P\simeq\frac{1}{3}\left(\frac{\pi^{2}}{15}\frac{1}{\beta^{4}}\right)\left[1+\frac{g}{2}\left(1-\frac{15}{4\pi^{2}}\left(\beta m\right)^{2}\right)\right]{\text{ \ \ }}(\beta m\ll1)\,,\label{eq:PPodolskyBetaMSmall}
\end{equation}
\begin{equation}
\varepsilon\simeq\left(\frac{\pi^{2}}{15}\frac{1}{\beta^{4}}\right)\left[1+\frac{g}{2}\left(1-\frac{5}{4\pi^{2}}\left(\beta m\right)^{2}\right)\right]\qquad(\beta m\ll1)\,,\label{eq:EPodolskyBetaMSmall}
\end{equation}
which are similar to but not equal to Eqs. (\ref{eq:PProcaBetaMSmall},\ref{eq:EProcaBetaMSmall})
since they include the Maxwellian contribution to the terms coming
from the massive photon.

By inverting Eq.~(\ref{eq:EPodolskyBetaMSmall}) and substituting
the result in (\ref{eq:PPodolskyBetaMSmall}), one gets:
\begin{equation}
P\simeq\frac{\varepsilon}{3}\left(1-4\frac{M^{2}}{\sqrt{\varepsilon}}\right)\qquad(\beta m\ll1)\,,\label{eq:P(E)PodolskyBetaMSmall}
\end{equation}
if one defines
\begin{equation}
M^{2}=M^{2}(m)\equiv\frac{1}{6}\sqrt{\frac{15}{\pi^{2}}}\frac{\frac{g}{2}}{\sqrt{1+\frac{g}{2}}}\frac{m^{2}}{4} \, \, \, \, \, \, \, \, (\mbox{Bopp-Podolsky})\,.\label{eq:MPodolsky}
\end{equation}
Eq.~(\ref{eq:P(E)PodolskyBetaMSmall}) is formally the same as Eq.
(\ref{eq:P(E)ProcaBetaMSmall}), the difference being the definition
of parameter $M$: compare Eqs. (\ref{eq:MPodolsky}) and (\ref{eq:MProca})
keeping in mind that $m$ is very large in Bopp-Podolsky electrodynamics
while it is very small in Proca case. This fact guarantees that the
steps to calculate $a(t)$ for Bopp-Podolsky radiation are the same as
the ones previously followed in de Broglie-Proca's case, cf. sentences containing
Eqs. (\ref{eq:DiffEqE(a)})-(\ref{eq:a(t)ProcaBetaMSmall}). Therefore, 
the scale factor for a Bopp-Podolsky photon gas in the high-energy regime
is \footnote{In Eq. (\ref{eq:a(t)PodolskyPrimeval}), $a(t_{0})=a_{0}$ can not
be interpreted as the value of the scale factor today because $a(t)$
is valid for the early universe. }:
\begin{equation}
a(t)\simeq a_{0}\left\{ 1+2H_{0}(t-t_{0})\left[1+\frac{M^{2}}{\sqrt{\varepsilon_{0}}}H_{0}(t-t_{0})\right]\right\} ^{1/2}\label{eq:a(t)PodolskyPrimeval}
\end{equation}
$(\beta m\ll1)$, just like in de Broglie-Proca's future universe $-$ see Eq.~(\ref{eq:a(t)ProcaBetaMSmall})
and interpretation below it. Eq.~(\ref{eq:a(t)PodolskyPrimeval})
essentially means that Bopp-Podolsky massive photons may not produce sensible
effects in cosmic dynamics. This will be confirmed in the following
analysis of the non-approximate solution to Friedmann equations.

\begin{figure}
\begin{centering}
\includegraphics[height=6.4cm, width=8.5cm]{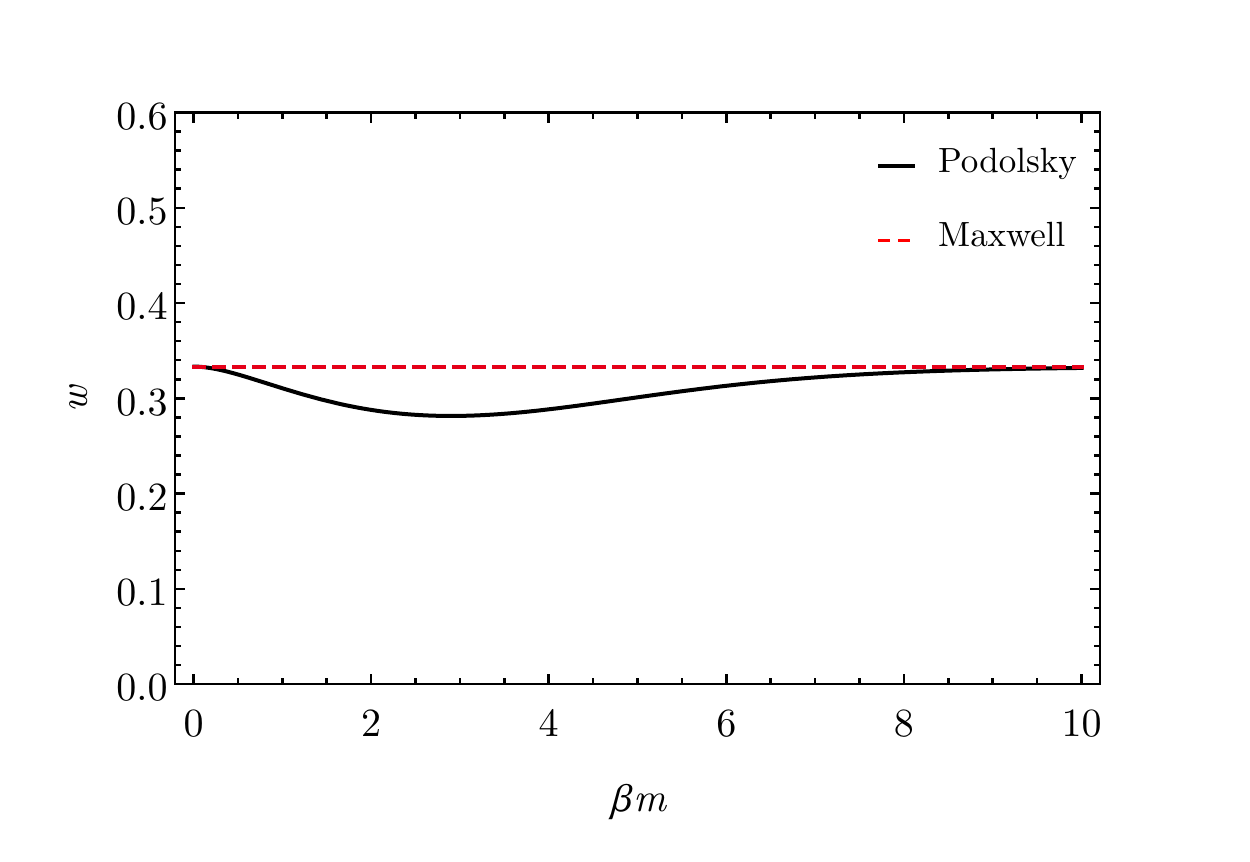}
\par\end{centering}
\caption{Plot of $w$ as a function of parameter $\beta m$. The dashed line
corresponds to $w=1/3$ as expected for a massless photon gas. The
continuous line exhibits the behavior of the EOS parameter $w$ for Bopp-Podolsky theory with $g=3$; in this case, $w$
is minimum for $\beta m$ equal to 2.899.}
\label{fig:w(BetaM)}
\end{figure}

The ratio of Eqs. (\ref{eq:PressurePodolsky}) and (\ref{eq:EnergyPodolsky})
lead to the parameter of the barotropic equation of state,
\begin{equation}
w=\frac{P}{\varepsilon}=\frac{1}{3}\frac{1}{(1+f)}\,,\label{eq:w}
\end{equation}
where
\begin{equation}
f=f(\beta m)=\frac{\frac{g}{30}(\beta m)^{4}\sum_{k=1}^{\infty}\frac{K_{1}(k\beta m)}{k\beta m}}{1+\frac{g}{10}(\beta m)^{4}\sum_{k=1}^{\infty}\frac{K_{2}(k\beta m)}{(k\beta m)^{2}}}\label{eq:f}
\end{equation}
is the function distinguishing the Maxwellian result ($P=\varepsilon/3$;
$f=0$) from Bopp-Podolsky electrodynamics. Fig. \ref{fig:w(BetaM)} shows
the plot for $w=w(\beta m)$.

%%% Figure %%%

%%%%%%%%%%

The Maxwell equation of state parameter $w=w_{M}=1/3$ is recovered for
both $\beta m\rightarrow0$ (i.e., $\beta m\ll1$) and $\beta m\gg1$; this means that the presence of the massive photon can not sensitively alter cosmic dynamics either in the distant past or in the present/future. In the distant past ($\beta m\ll1$) the mean thermal energy of the universe is much greater than the rest energy of the massive photon, and the Bopp-Podolsky photon behaves as an ultra-relativistic particle. In
the other limit, the condition $\beta m\gg1$ is satisfied whenever the
photon mass is much greater than the temperature $T=\beta^{-1}$;
this is a condition fulfilled by the present-day universe
whose temperature is $T_{0}\simeq2.73\text{K}\simeq2.35\times10^{-4}\text{eV}$
while $m$ is $3.70\times10^{10}\text{eV}$ at least $-$ Ref.~\cite{Bufalo2012}.
In spite of the equivalence Maxwell-Podolsky concerning the parameter
$w$ in the limits $\beta m\ll1$ and $\beta m\gg1$, it is worth
mentioning that this is not the case for the energy density: $\varepsilon\rightarrow\varepsilon_{M}$
when $\beta m\gg1$, but $\varepsilon\rightarrow(5/2)\varepsilon_{M}$
if $\beta m\rightarrow0$, cf. Fig.~\ref{fig:E(BetaM)}.

The maximum influence of Bopp-Podolsky massive photons to the equation of state
corresponds to the minimum of the curve $w(\beta m)$ in Fig. \ref{fig:w(BetaM)}:
$w_{\text{min}}\simeq0.282$ for $\beta m\simeq2.899$, when the mass
is about three times the value of the mean thermal energy of the universe.
At this value of $\beta m$,
\begin{equation}
\frac{\Delta w}{w_{M}}=\frac{w_{M}-w_{\text{min}}}{w_{\text{min}}}\simeq15.4\% \nonumber 
\end{equation}
and the universe attains its minimum deceleration compared to the
one achieved by a massless photon gas. This is true once $\ddot{a}/a\propto(1+3w)$,
as one can easily show from Eqs. (\ref{eq:FriedEq}), (\ref{eq:FluidEq})
and $P=w\varepsilon$.

%%%%%%%%%%%%%%%%%%%%%%%%%%%%%%%%%%%%%%%%%%%

\section{Final Remarks\label{sec:final-remarks}}

This paper analyzes the effects that a massive photon accommodated
by de Broglie-Proca and Bopp-Podolsky theories could produce on cosmic dynamics. The approach is based on the hypothesis of thermal equilibrium
which allows the construction of an equation of state for the massive photon
gas in each case. It was shown that a barotropic equation
of state $P\neq\varepsilon/3$ is produced; this is true for both
de Broglie-Proca and Bopp-Podolsky electrodynamics. 
(This is different from what happens for
non-linear electrodynamics in a background field, where $P=\varepsilon/3$
is preserved \cite{Akmansoy2014} and no new cosmological phenomenon
appears.) However, the departure of the EOS from a Maxwellian form
does not guarantee a significant modification in the functional form
of the scale factor $a\propto t^{1/2}$ typical of massless radiation.

In particular, the effect of a \textit{de Broglie-Proca} photon mass is completely
negligible for cosmic dynamics when one considers the more realistic
context where dark matter and dark energy are present. In fact, as
shown in Sect. ``de Broglie-Proca cosmology'', from the early universe
until a future where $a_{\text{future}}=10^{12}a_{0}$, de Broglie-Proca's radiation
behaves approximately as Maxwell's: $\varepsilon\sim a^{-4}$. In
addition, observations \cite{Planck} show that the energy density
of radiation ($\varepsilon_{\gamma}$) in the present-day universe
is ten thousand times smaller then the matter energy density ($\varepsilon_{m}$)
today, i.e., $\varepsilon_{\gamma0}\sim10^{-4}\varepsilon_{m0}$, regardless
of the nature of the cosmic photon gas (either massive or massless).
Thus, in a future where the scale factor amounts to $a_{\text{future}}$,
one estimates $\varepsilon_{\gamma,\text{future}}\simeq10^{-16}\varepsilon_{m,\text{future}}$
because $\varepsilon_{m}\sim a^{-3}$: this makes radiation dynamically
irrelevant in the face of matter.

If one insists on advancing even further towards the future, considering
$a>a_{\text{future}}$, the de Broglie-Proca mass begins to take its toll; $\varepsilon_{\gamma}$
slowly modifies its functional dependence on the scale factor, evolving
from $a^{-4}$ to $a^{-3}$ in the future infinity. In this regime
($a\gg10^{14}a_{0}$) radiation behaves as non-relativistic, but with
an initial condition where the radiation energy density is 16 orders
of magnitude smaller than matter energy density. Consequently, matter
utterly dominates radiation. The situation is deeply aggravated in
the presence of some type of dark energy (DE) scaling as $\varepsilon_{\text{DE}}\sim a^{-n}$
where $n<2$; then $\varepsilon_{\gamma,\text{future}}\ll10^{-14}\varepsilon_{\text{DE,future}}$
rendering the de Broglie-Proca mass even more negligible compared to the dark component.

As seen in Sect. ``Bopp-Podolsky cosmology'', \textit{Bopp-Podolsky}
electrodynamics differs from de Broglie-Proca's in two fundamental ways: the
mass of the photon is humongous (instead of been extremely small) and
there are derivative terms in the field strength entering the Lagrangian
(instead of quadratic terms involving $A^{\mu}$). Someone will argue
that these derivative terms lead to the appearance of ghosts, that
a theory with such a plague should be immediately discarded as inconsistent.
However, some works analyze this issue \textendash{} e.g., Ref. \cite{Kaparulin}
\textendash{} and they point to a well-behaved type of ghosts. In
fact, Ref. \cite{Kaparulin} shows that Bopp-Podolsky electrodynamics belongs
to a wide class of higher-derivative systems admitting a bounded integral
of motion which makes them dynamically stable despite their canonical
energy being unbounded. Thermodynamics of Bopp-Podolsky massive photon gas does affect cosmic dynamics,
and this occurs for $0\leq\beta m\lesssim8$ (see Fig. \ref{fig:w(BetaM)}).
However, this influence is not pronounced: The massive term is not able to produce
any sensible deviation of cosmic dynamics from a massless photon gas
in the radiation-dominated era. In particular, Bopp-Podolsky radiation
can not produce an accelerated expansion in the early universe since
its EOS parameter respects: $0.282<w<1/3$.

This paper shows that the maximum influence of Bopp-Podolsky theory on
cosmic dynamics takes place for $\beta m\simeq2.899$. If one chooses
the minimum value $m=37\text{GeV}$ in accordance with Eq. (\ref{eq:PodolskyMass}),
this corresponds to $kT\simeq13\text{GeV}$; i.e., one order of magnitude below the energy scale of electro-weak unification. Notice that the cosmic dynamics for the Bopp-Podolsky radiation was determined at
all times in terms of the product $\beta m$: it does not depend directly
on the photon mass. In this sense, our work implies that the standard cosmological model
does not rule out Bopp-Podolsky massive photon gas as a real possibility.
This very fact, along with the success of predictions by 
generalized quantum electrodynamics \cite{Bufalo2011,Bufalo2013,Bufalo2014,Bufalo2012},
motivates the continuing study of Bopp-Podolsky theory. In addition, the
massive mode of Bopp-Podolsky photon may interact with charged particles
present in the cosmic soup \footnote{In Stueckelberg theory the massive photon does interact: there is
a coupling with neutrinos and charged leptons \cite{Ruegg}.}. The resulting dynamics of this interaction is not trivial and a realistic
model should take it into account; this might be a suitable subject
for future investigation.

\bigskip

%%%%%%%%%%%%%%%%%%%%%%%%%%%%%%%%%%%%%%%%%%%

\begin{acknowledgments}
RRC is grateful to Prof. R. Brandenberger and Bryce Cyr at McGill Physics Department. EMM and CNS thank CAPES/UNIFAL-MG (Brazil) for financial support. LGM (grant 112861/2015-6) and BMP acknowledge CNPq (Brazil) for partial financial support.
\end{acknowledgments}

%%%%%%%%%%%%%%%%%%%%%%%%%%%%%%%%%%%%%%%%%%%%

\end{document}